\documentclass[fleqn,usenatbib]{mnras}

\usepackage{newtxtext,newtxmath}
\usepackage[T1]{fontenc}
\usepackage{ae,aecompl}

\usepackage{siunitx}
\usepackage{placeins}
\usepackage{pdflscape}
\usepackage[super]{nth}

\usepackage{graphicx}	% Including figure files
\usepackage{amsmath}	% Advanced maths commands
\usepackage{amssymb}	% Extra maths symbols

\newcommand{\tripleseven}{O I $7774\,\si{\angstrom}$}
\newcommand{\halpha}{H$\alpha$}
\newcommand{\novacar}{ASASSN-18fv}
\newcommand{\solarmass}{M_{\odot}}

\DeclareSIUnit\day{d}
\DeclareSIUnit\minute{min}
\DeclareSIUnit\hour{hr}

\title[Fe-rich circumbinary disc around a nova]{Classical Nova Carinae 2018: Discovery of circumbinary iron and oxygen}

\author[D. McLoughlin et al.]{
Dominic McLoughlin,$^{1}$\thanks{E-mail: dominic.mcloughlin@physics.ox.ac.uk (DM)}
Katherine M. Blundell$^{1}$ and
Steven Lee $^{2, 3}$
\\
$^{1}$Physics Department, University of Oxford, Keble Rd, Oxford OX1 3RH, United Kingdom\\
$^{2}$Research School of Astronomy and Astrophysics, Australian National University, Canberra, ACT 2611\\
$^{3}$Anglo-Australian Telescope, Coonabarabran, NSW 2357, Australia\\
}

\date{Accepted XXX. Received YYY; in original form ZZZ}

\pubyear{2019}

\begin{document}
\label{firstpage}
\pagerange{\pageref{firstpage}--\pageref{lastpage}}
\maketitle

\begin{abstract}
We present time-lapse spectroscopy of a classical nova explosion commencing 9 days after discovery. These data reveal the appearance of a transient feature in \ion{Fe}{ii} and [\ion{O}{i}]. We explore different models for this feature and conclude that it is best explained by a circumbinary disc shock-heated following the classical nova event. Circumbinary discs may play an important role in novae in accounting for the absorption systems known as THEA, the transfer of angular momentum, and the possible triggering of the nova event itself.
\end{abstract}

\begin{keywords}
novae -- shock waves -- stars: individual: ASASSN-18fv
\end{keywords}

%%%%%%%%%%%%%%%%%%%%%%%%%%%%%%%%%%%%%%%%%%%%%%%%%%

%%%%%%%%%%%%%%%%% BODY OF PAPER %%%%%%%%%%%%%%%%%%

\section{Introduction}
A classical nova is an explosion on the surface of a white dwarf and a less-evolved companion in a close binary. The companion is usually a main sequence star, or a red giant. Hydrogen gas accretes from the companion onto the surface of the white dwarf via an accretion disc (in the absence of strong magnetic fields), until the high temperatures and pressures at the bottom of the accreted layer pass a critical threshold and the hydrogen gas commences nuclear fusion. This powers a very rapid thermonuclear runaway reaction, and the resultant explosion ejects a shell, on average of mass $\num{2e-4}\,M_{\odot}$ \citep{Gehrz1998}.

Classical nova explosions are rare opportunities to study what happens when a binary star system consisting of a white dwarf and a donor star undergo such a perturbation - dynamically, thermally and chemically.  Spectroscopic study of the immediate post-explosive behaviour following a nova is a powerful tool to reveal, via the Doppler effect, dynamical changes, and via the appearance and subsequent disappearance of various emission lines, the changes that arise following nucleosynthesis and shock heating. 

In March 2018, a nova detonated in Carina, a circumpolar location in the Southern Hemisphere sky, very convenient for continuous observation. It was discovered by the All Sky Automated Survey for SuperNovae\footnote{http://www.astronomy.ohio-state.edu/$\sim$assassin/} (hereafter ASASSN) and hence was named as \novacar\ as well as Nova Carinae 2018, and V906 Carinae. We followed this up spectroscopically with dense coverage (an average of 13.8\,hr per nychthemeron) from days 9-20, with sparse coverage after that. Much rich behaviour was observed giving insights into different aspects of nova evolution in the first few weeks after explosion. In this paper we focus on remarkable transient spectral features associated with \ion{Fe}{ii} and [\ion{O}{i}] lines.

As detailed in \citet{Arai2016}, there has been much debate surrounding the origin of multiple absorption systems in classical novae. These systems were initially identified by \citet{McLaughlin1950} as the `principal' and `diffuse-enhanced' absorption systems. Since then, \citet{Williams2008} has suggested another type of absorption component, the transient heavy element absorption (THEA). They suggest a reservoir of circumbinary gas is the origin of the low radial velocity THEA system. Circumbinary discs around cataclysmic variables were theoretically investigated by \citet{Spruit2000}, who believe such discs could be important in explaining angular momentum transfer in classical novae. However, to our knowledge this is the first direct observation of a circumbinary disc surrounding a classical nova in the literature to date, although \citet{Monnier2006} does hypothesise the existence of a circumbinary reservoir around the recurrent nova RS Ophiuchi.

In Section \ref{sec:observations}, we describe our round-the-clock spectroscopic coverage of the nova \novacar. In Section \ref{sec:interpretation}, we confront our data with a circumbinary disc model. In Section \ref{sec:disc_type}, we describe the context and theoretical background for interpreting our observations. In Section \ref{sec:alternatives}, we discuss alternative possible models including a pinwheel geometry as reported by \citet{Tuthill2007}. In Section \ref{sec:conclusions} we present our conclusions.

\section{Observations} \label{sec:observations}
%  Conventions
The earliest detection of the nova \novacar\ was at 2458193.819444\,JD by the Evryscope-South observatory, according to \cite{Corbett2018}, which we refer to as the discovery date for the system. It was independently discovered after this by \citet{Atel11454} using AS-ASSN. In this paper, we reference times as days since commencement of brightening (2458193.819444\,JD). This will be important for our timing data, which is given in this format unless indicated otherwise.
We use the unit mAU for milli-Astronomical Units as this is a convenient length scale to discuss the orbital separations and circumbinary radii of such a classical nova system.

\subsection{Global Jet Watch observatories}
Spectra of \novacar\ spanning a wavelength range of approximately $5800\,\si{\angstrom}$ to $8400\,\si{\angstrom}$ and with a spectral resolution
of $R\,{\sim}\,4000$ were observed starting from +9.514d. These were carried out with the multi-longitude Global Jet Watch telescopes, each of which is equipped with an Aquila spectrograph; the design and testing of these high-throughput spectrographs are described by S. Lee et al. (2020, in preparation). The observatories, astronomical operations, processing, and calibration of the spectroscopic data streams are described in K. M. Blundell et al. (2020, in preparation).

We used a range of exposure times throughout the observations of \novacar\ to ensure that we could both capture unsaturated Balmer \halpha\ in the shorter exposure times, and detect weaker signals in the longer exposure times. We also varied the exposure times as the light curve changed, as classical novae change luminosity by several orders of magnitude. A detailed breakdown of exposure times is given in Table \ref{tab:exposure_times}.

\begin{table}
 \caption{List of exposure times for observations of \novacar, in decreasing order of number of observations, as of \nth{4} February 2020.  The exposure times were chosen to suit the instantaneous brightness of the target, for example to avoid saturating the H-$\alpha$ line.}
 \label{tab:exposure_times}
 \begin{tabular}{cc}
  \hline
  Exposure time & Number of spectra \\
  \hline
  100 & 1370 \\[2pt] 
  300 & 1194 \\[2pt]
  200 & 217 \\[2pt]
  150 & 140 \\[2pt]
  10 & 110 \\[2pt]
  60 & 64 \\[2pt]
  250 & 52 \\[2pt]
  60 & 64 \\[2pt]
  3 & 29 \\[2pt]
  3000 & 6 \\[2pt]
  1000 & 3 \\[2pt]

 \end{tabular}
\end{table}

\subsection{Epochs of observations}
\begin{figure}
	\includegraphics[width=\columnwidth]{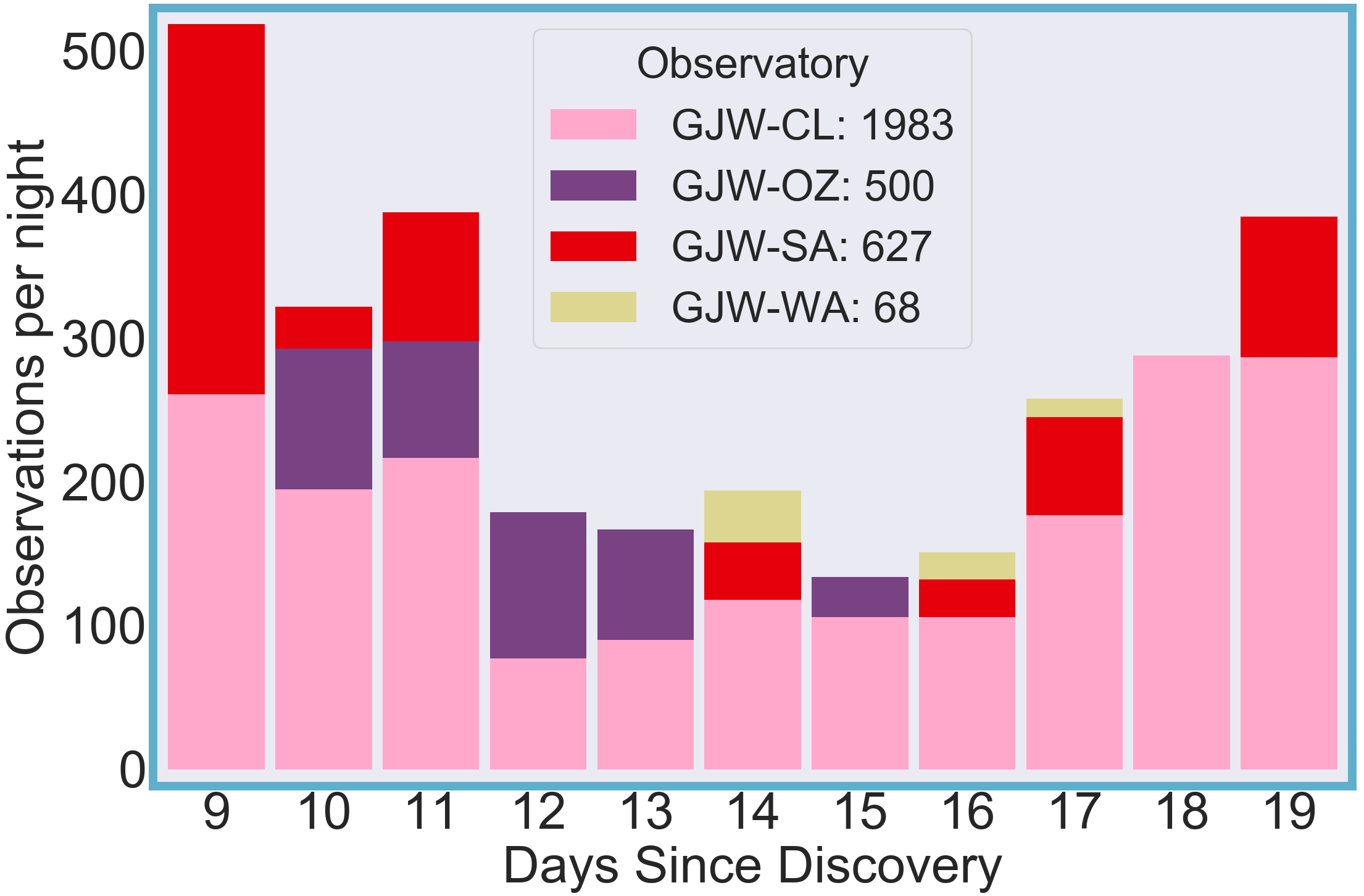}
    \caption{Stacked histogram showing the number of spectra taken each day per observatory during the early days of observation. CL - Chile, OZ - East Australia, SA - South Africa, WA - West Australia.}
    \label{fig:nova_car_observations_histogram}
\end{figure}

Figure \ref{fig:nova_car_observations_histogram}  is a histogram of the number of spectroscopic observations captured, binned daily for the first ten days of observations. The different colours represent the different observatories. What is notable about this dataset is that we have an unbroken run of over 100 spectra per night for the fastest evolving part of the light curve - these ten days. Furthermore, the first three days have over 300 spectra each. We have $22.5\,\si{\hour}$ of CCD exposure time in the first $24\,\si{\hour}$ of the observation set. This is made possible because the telescopes are separated in longitude around the Earth, such that when one is in daylight, another is in night.  In total 3256 spectra were taken up to 2018 Sep 30, extending over a timespan of 6\,months since the detonation. Our investigations of other spectral features and their evolution will be presented in subsequent papers.

\subsection{Spectral features}
Figure \ref{fig:stacked_by_days} illustrates a spectral feature, somewhat in the shape of the silhouette of a capital M with sloping outer edges, which we noticed in several of our spectra while investigating other aspects of the evolving nova. This feature is most prominent from +13.5d to +20d, shown by the central column in Figure \ref{fig:stacked_by_days}. Its origins can be traced as early as day +13.1d (top left panel), while traces of this structure remain until +50d (bottom right panel), but we note that it does not constitute a detection at these extremes of times as the signal-to-noise is too low. Each of five different colours represents an overlaid spectral feature corresponding to a different rest wavelength of an \ion{Fe}{ii} or [\ion{O}{i}] line. We see no evidence for any spectral shape resembling this feature in either \halpha\ or \ion{He}. It is remarkable that the sloping outer boundaries of this feature overlay one another with such good agreement for each of the five lines. The modal velocities correspond to the velocities of the outermost particles in a disc, per \cite{Elitzur2012}. This signal is consistent with the nova having very low ($< 50\,\si{\km\per\second}$) systemic velocity. This is corroborated by the pre-maximum spectra, which show narrow absorption lines with Doppler shifts roughly $\Delta \lambda < 1\,\si{\angstrom}$ relative to the rest wavelengths, within the spectral resolution.

We see no significant change in the velocity width of the spectral features shown in Figure \ref{fig:stacked_by_days}. We discuss the implications of this observation in Section \ref{sec:spectral_shape} below.

\subsection{Iron and Oxygen disc detection}
\begin{figure}
	\includegraphics[width=\textwidth]{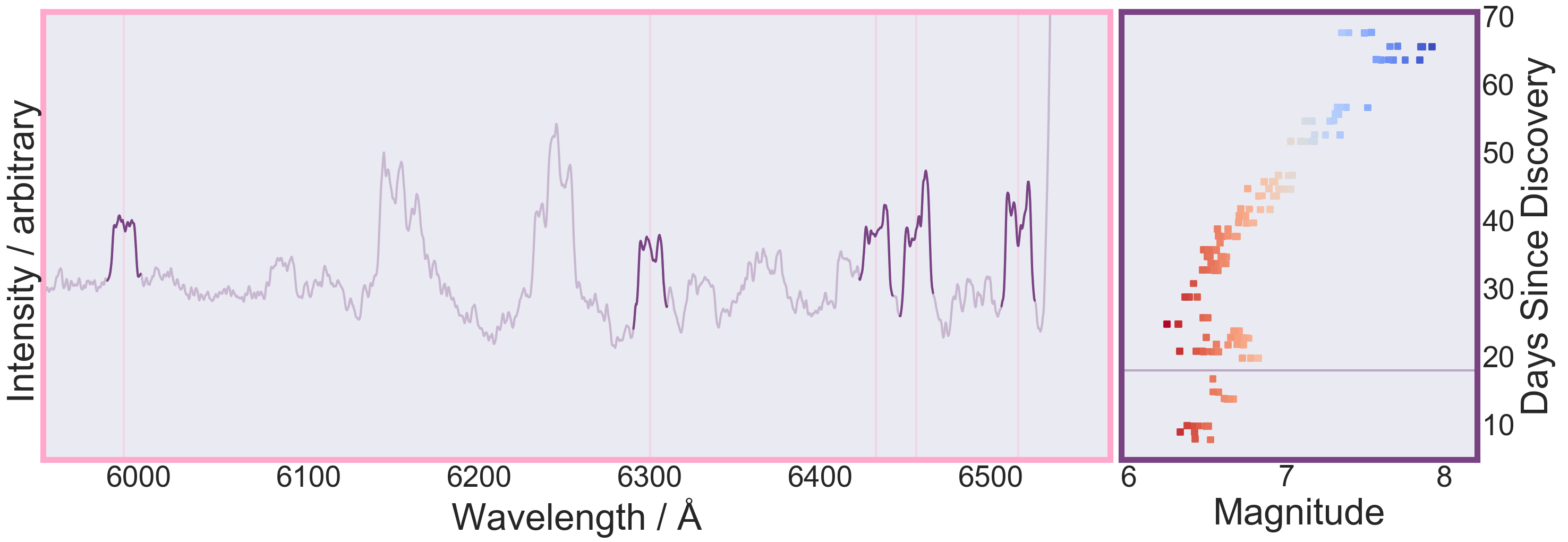}
    \caption{A single spectrum of exposure time 150\,s observed at the GJW-CL observatory at $+17.905$\,d, showing the characteristic signatures of the disc feature, at a number of different wavelengths (identified as four different \ion{Fe}{ii} transitions at $5991\,\si{\angstrom}$, $6432\,\si{\angstrom}$, $6456\,\si{\angstrom}$, $6516\,\si{\angstrom}$, and the forbidden [\ion{O}{i}] $6300\,\si{\angstrom}$ transition) indicated in Fig\,\ref{fig:stacked_by_days}.}
    \label{fig:demo_spectrum}
\end{figure}

We observe transient disc-like emission profiles at five different central wavelengths across our spectra, highlighted in bold in Figures \ref{fig:demo_spectrum} and \ref{fig:full_spectra_demos}. It is clear from Figure \ref{fig:stacked_by_days} that aligning the profiles on rest wavelengths of known \ion{Fe}{ii} emission lines ($5991\,\si{\angstrom}$, $6432\,\si{\angstrom}$, $6456\,\si{\angstrom}$, $6516\,\si{\angstrom}$) and the $6300\,\si{\angstrom}$ [O I] line gives remarkable concordance in the velocity profiles at +17.8d. The modal Doppler shift velocity, most naturally interpreted as the rotational velocity at the outer edge of a disc, is measured at $-266 \,\si{\km\per\second}$ for the blue wing, and $254 \,\si{\km\per\second}$ for the red wing. This is the same ballpark speed as the initial low velocity component absorption features observed in \halpha\ and \tripleseven\ lines (D. McLoughlin et al. 2020 in preparation). The M-shaped profile is unambiguously present from +13.9d to +19.8d, with marginal hints outside this range.

\subsection{Profile variations}
\begin{figure}
	\includegraphics[width=\textwidth]{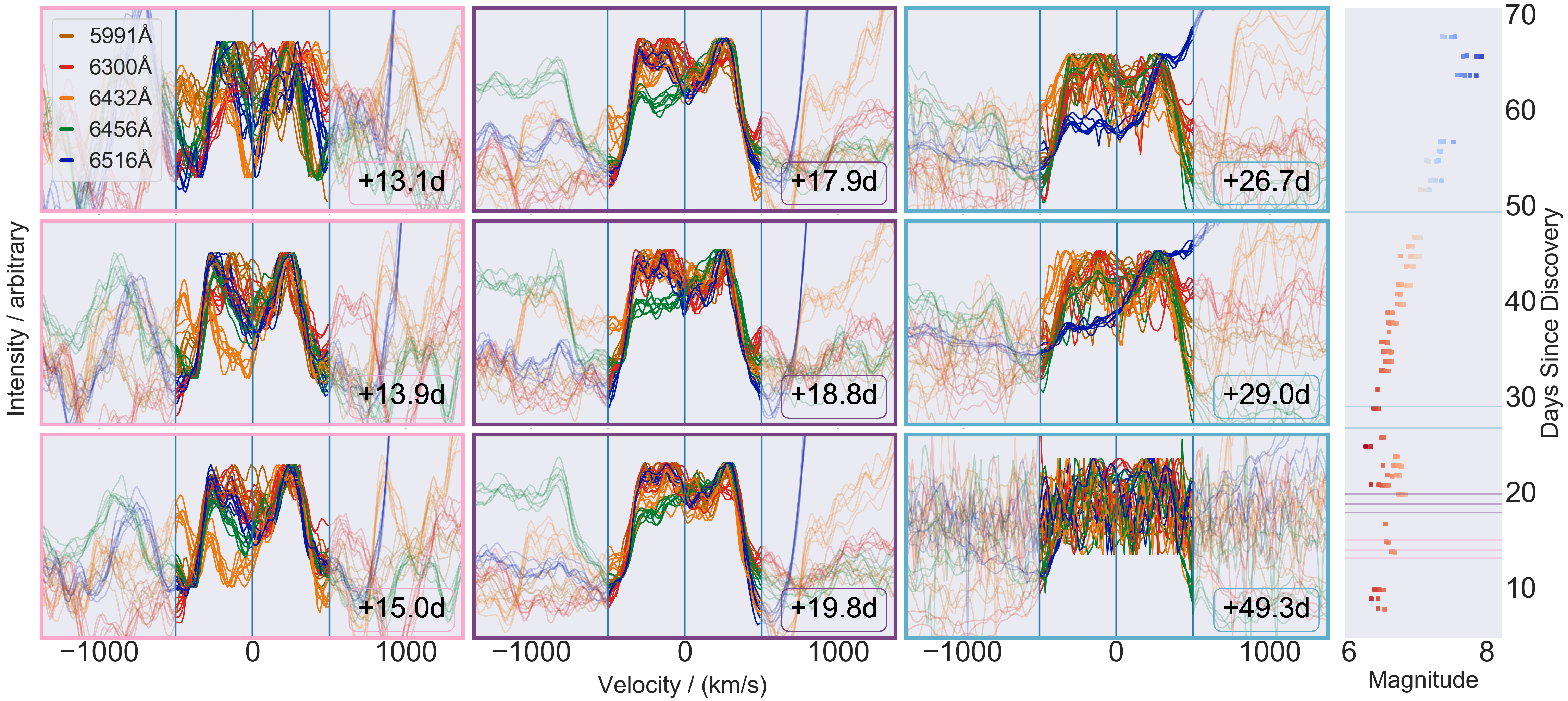}
    \caption{The central column shows a clear sloping M-shaped spectral feature in four lines of \ion{Fe}{ii} and one line of [\ion{O}{i}]. Spectra taken at nine different epochs, shown in velocity space using the rest wavelength of each line as the zero as zero radial velocity.  The rightmost panel shows the light curve, with time increasing up the page, and magnitude increasing to the right. The observation times are shown superimposed on the light curve as horizontal green, peach and lilac lines. After at best a marginal detection at +13.1d, we see the \ion{Fe}{ii} and [\ion{O}{i}] signature emerge from +13.9d, and last until +19.8d. At each epoch, we superimpose six spectra taken within minutes of each other to demonstrate that this is a persistent signal, not just spectral noise.}
    \label{fig:stacked_by_days}
\end{figure}

\begin{figure}
	\includegraphics[width=\textwidth,angle=0]{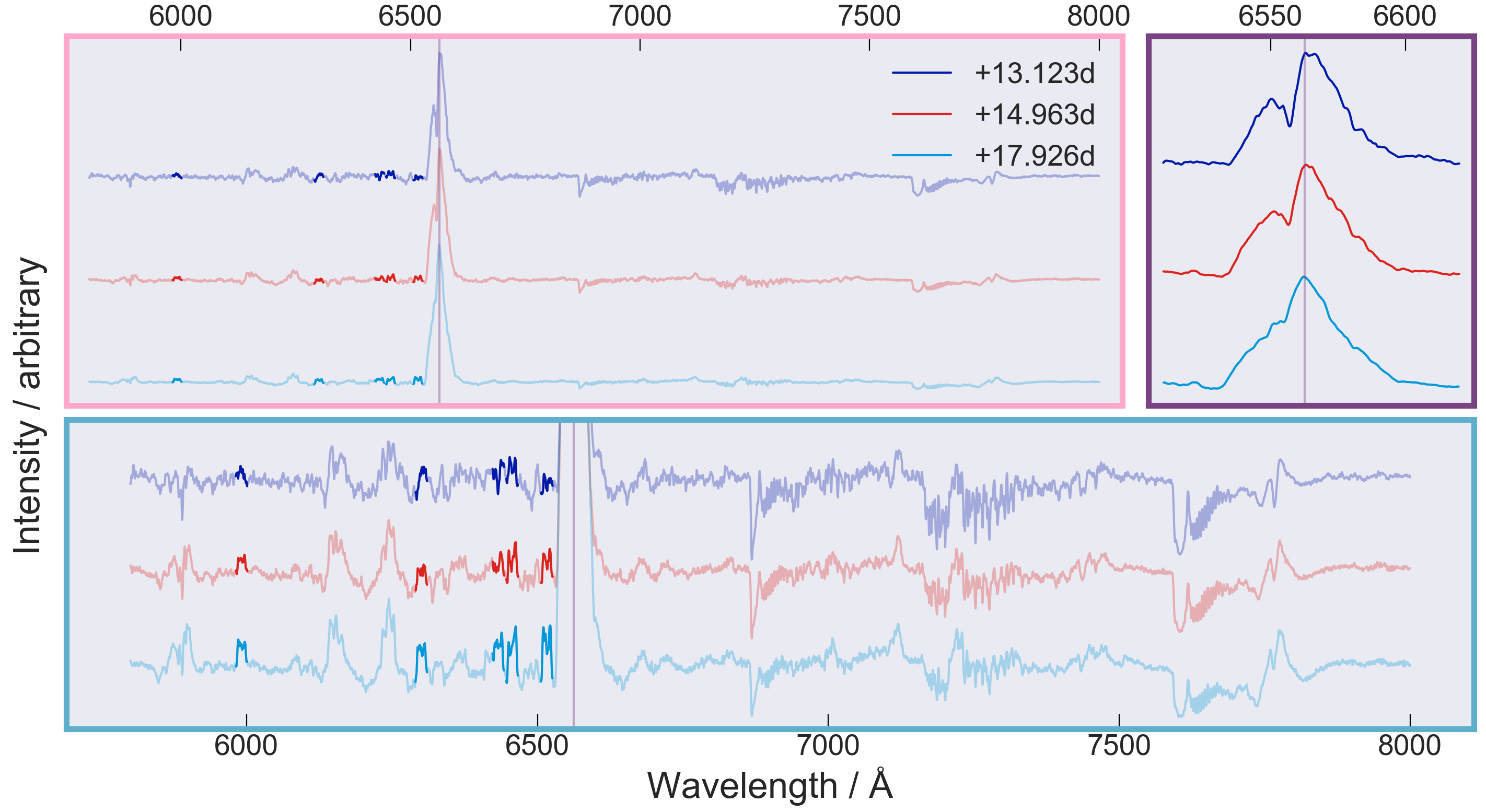}
    \caption{Three example spectra taken at +13.123d (dark blue), +14.963d (red), +17.926d (light blue). The three panels clockwise from top left show the full spectra, a close-up of the \halpha\ complex, and the full spectra cropped to show the spectral features other than \halpha. The bold segments of these spectra are the same sections as those shown in bold in Figure \ref{fig:stacked_by_days}.}
    \label{fig:full_spectra_demos}
\end{figure}

For each epoch shown in Figure \ref{fig:stacked_by_days}, we superimpose six spectra taken within minutes of each other, to demonstrate that this is a persistent signal, not simply spectral noise. While all these lines show the overall M-shaped profile, there is some variation between the profiles of different spectral lines within each epoch. We investigated whether telluric absorption lines had affected some of the \ion{Fe}{ii} and [\ion{O}{i}] profiles, and established that while there are some such lines, they were too narrow and weak to significantly alter the profiles in our spectra.

There is a notable deviation from the clear M-shape seen in Figure \ref{fig:stacked_by_days} at early times in the $6432\,\si{\angstrom}$ line (yellow, left column), and at later times the $6456\,\si{\angstrom}$ line (green, middle column). For each of these two cases, the emission decrement relative to the other lines is limited to the blue-shifted peak. The decrement itself evolves with time, disappearing for some of the spectra and returning later in others, which suggests that a real physical evolution is unfolding, which merits a detailed numerical simulation, in a future paper.

\onecolumn

\twocolumn
\section{Interpretation}\label{sec:interpretation}
\subsection{Spectral shape}\label{sec:spectral_shape}
The double-peaked spectral features (showing such structure for many \ion{Fe}{ii} lines and for the forbidden [\ion{O}{i}] line (very different transitions) with such similar boundaries in velocity) are naturally interpreted as rotating gas, as evinced by Fig\,\ref{fig:best_fit_disc_uniform_density}. We discuss other possible interpretations in detail in Section \ref{sec:alternatives}, but we find that rotating gas best fits the data. We fit this model to our data in Section \ref{sec:fitting_disc}. 

\subsection{Disc model and fitting} \label{sec:fitting_disc}
Following \cite{Elitzur2012}, we model the nova as having a kinematically double-peaked Keplerian ring. Our simple model considers a flat disc viewed edge-on, with all particles on Keplerian orbits. We assume constant density throughout the ring for simplicity. The double-peaks arise naturally from the geometry; considering contours of constant line-of-sight velocity, the longest such contours correspond to the modal velocity, or the peaks of the spectral signature. The contour of zero radial velocity is caused by material moving perpendicular to the line-of-sight. This is a shorter contour than that for higher absolute Doppler shifts, explaining the dip between the peaks. 

\begin{figure}
	\includegraphics[width=\columnwidth]{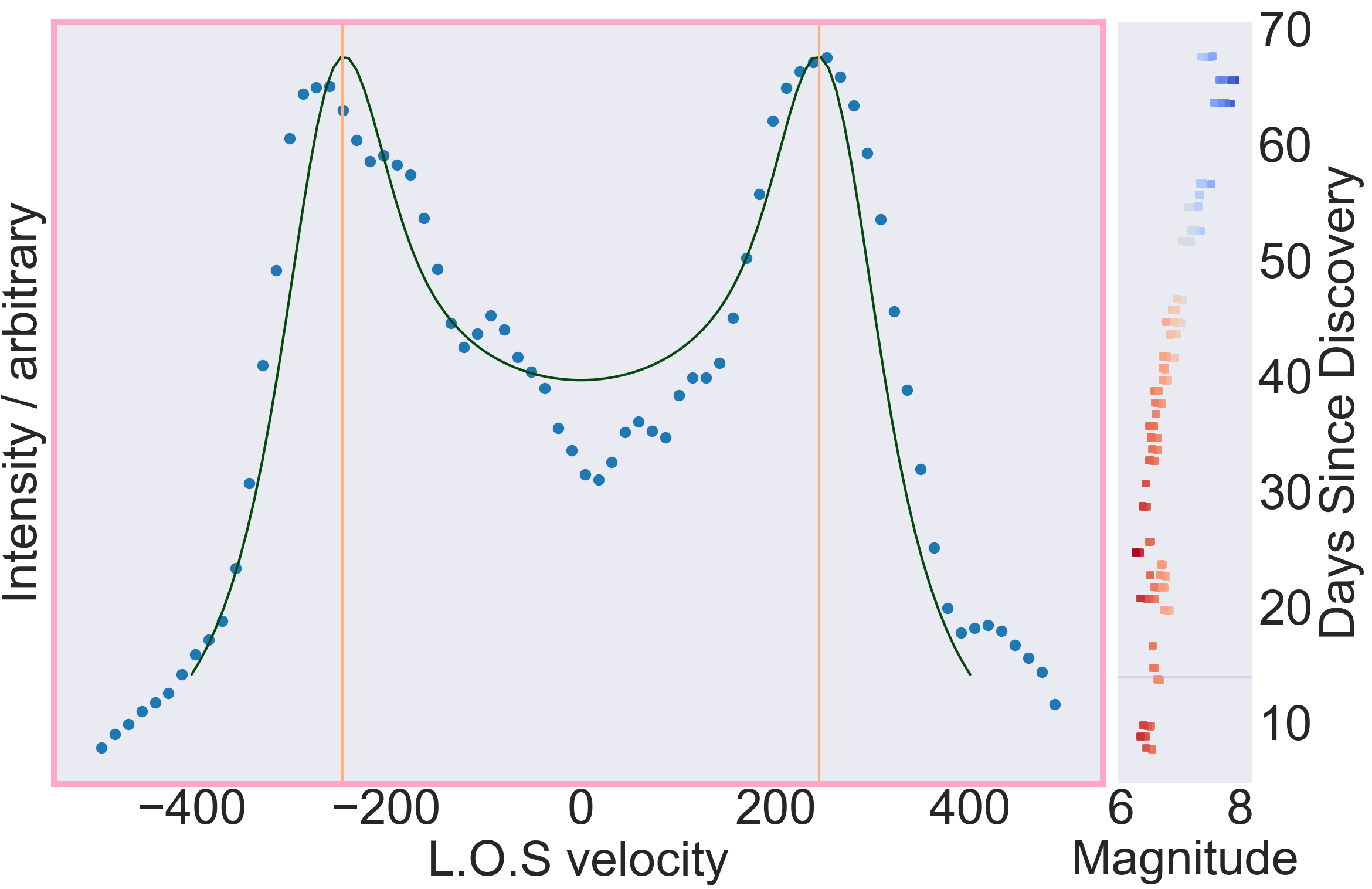}
    \caption{Best fit for a rotating Keplerian disc with uniform surface density, viewed edge-on. Data are taken from the $6516\,\si{\angstrom}$ \ion{Fe}{ii} line signal, taken at +13.928d at the GJW-CL observatory in Chile, indicated on the light curve with a horizontal line. The bins are $0.3\,\si{\angstrom}$ wide, or $14\,\si{\km\per\second}$ at this wavelength. The peach-coloured vertical lines indicate the velocity of the fit at the outer edge of the disc.}
    \label{fig:best_fit_disc_uniform_density}
\end{figure}

We fit this model to our data as shown in Figure \ref{fig:best_fit_disc_uniform_density}, using three free parameters: (i) total Keplerian central mass / radius of emissive material ($M_{\textrm{tot}}/r_{\textrm{disc}}$) (ii) radius ratio of disc ($r_{\textrm{in}}/r_{\textrm{out}}$) (iii) characteristic velocity ratio ($v_{\textrm{rot}}$/$v_{\textrm{thermal}}$). Table \ref{table:mcmc_fitting_params} details the upper and lower bounds we used. The mass bounds were designed to incorporate the full range of what could be expected for this system. Considering that the likely mass range for the white dwarf in a classical nova system is at least restricted to $0.6\,M_{\odot}$ to $M_{\textrm{CH}}^{} = 1.4\,M_{\odot}$ (the Chandrasekhar limit)\citep{Shara2018}, we believe the combined mass of the two stars to be contained within the interval $0.5\,M_{\odot} < M_{\textrm{tot}} < 3\,M_{\odot}$. The inner/outer radius was allowed to vary from 0, the disc, to 1, the vanishingly thin ring. The relative velocity scales was allowed to range from 0 to 20, as per \citet{Elitzur2012}. 

The results of the fitting are in Table \ref{table:best_fit_disc_uniform_density}. We utilised an MCMC fitting procedure based on the {\tt emcee} python package \citet{Foreman-Mackey2013}. Since the mass and outer radius are degenerate, we list the system parameters derived from the combined mass/outer radius free parameter for the case of the inner binary having $M_{\textrm{tot}} = M_{\odot}$.

\begin{table}
    \centering
    \caption{Fitting constraints for the uniform disc model using integration over disc area. Note that we combine the mass and outer radius parameters into one free parameter for the fit, since M/r is degenerate in this Keplerian model.}
    \label{table:mcmc_fitting_params}
    \begin{tabular}{c|c|c}
        \textbf{Parameter}      & \textbf{Lower bound} & \textbf{Upper bound}\\ \hline
        \textbf{Kepler Mass}           & 0.5\,$M_{\sun}$  &        3\, $M_{\sun}$         \\ \hline
        \textbf{Outer Radius}   & 0.1\,mAU  &   \num{e4}\,AU    \\ \hline
        \textbf{Radius ratio}   & 0        &    1 \\ \hline
        \textbf{v$_{\rm rot}$/v$_{\rm thermal}$}  & 0         &    20    \\ 
    \end{tabular}
\end{table}

\begin{table}
    \centering
    \caption{Fitted parameters for the uniform disc model using integration over disc area. Since total enclosed mass and radius are degenerate in this model, we present the values assuming the mass is $M_{\textrm{tot}} = M_{\sun}$}
    \label{table:best_fit_disc_uniform_density}
    \begin{tabular}{c|c}
        \textbf{Parameter}      & \textbf{Fitted value} \\ \hline
        \textbf{$M_{\textrm{tot}}$}           & 1$\,M_{\sun}$                \\ \hline
        \textbf{Inner Radius}   & 0.017\,mAU          \\ \hline
        \textbf{Outer Radius}   & 14.7\,mAU              \\ \hline
        \textbf{v$_{\rm rot}$/v$_{\rm thermal}$}  & 5.100                 \\ \hline
        \textbf{Outer Velocity} & $244\,\si{\km\per\second}$              
    \end{tabular}
\end{table}

\section{Circumbinary discs} \label{sec:disc_type} 
\subsection{Theoretical predictions}
According to the n-body, test particle, 3D simulations of \cite{Doolin2011}, there are 3 families of remarkably stable circumbinary orbits around binary star systems throughout eccentricity/mass ratio parameter space. Stable here means not that the material is confined to closed orbits whose trajectory through space is unchanging through time, but rather orbits that stably precess about the binary star system indefinitely. They reported that the rate of precession depends on both eccentricity and mass ratio in the sense that for a more eccentric system, or for a more skew mass ratio, the shorter the period of precession.

The distance from the binary centre of mass to the innermost stable radius for test particle orbits has been studied numerically and analytically \citep{Szebehely1980,Holman1999}. It is given as $r_c \approx f(q)a_B = f(q)r(1+q)$ where $a_B$ is the binary separation, q is the mass ratio m/M, and r is the orbital radius of the less massive component. $f(0.1) = 2$ and $f(0.3) \approx 2.3$

\subsection{Significance of timing} \label{sec:significance_timing}
As shown in Figure \ref{fig:stacked_by_days}, the M-shaped signal becomes clear at +13.9d. The signal then fades into noise after a week, with very little trace of the detailed shape remaining after +19.8d. Taking the speed of sound in the intervening medium between the explosion and a hypothetical circumbinary disc to be $1\,\si{\km\per\second}$, and the circumbinary radius to be two times bigger than the semimajor axis to be approximately $10\,$mAU, the time for a shockwave to reach a circumbinary disc would be around $17\,$d. This is the same ballpark time that it takes for the signal we observe to appear. 

What about the switch-off time? The signal appears to melt away ${\sim}10$\,days after appearing. This could be due to cooling of a shock-heated circumbinary disc. Under this interpretation, a shockwave travels radially outwards from the initial event for 14 days, then arrives at the disc and shock-heats it, rapidly causing emission in a profile that informs us of the underlying pre-existing circumbinary disc geometry at +13.9d. This recently heated disc is visible clearly for a week, but loses energy through radiation, and cools, which we observe as the disc signal grows fainter and fainter over the following week (from +19.8d to +27d). Crucially, in any model involving a sustained wind or ejection, it isn't trivial to conjure up a reason for the ejection to turn off in a week-long decay.

We also observe a change in dynamics over the week of strong signal. Close inspection of the spectrum at +17.8d in Figure \ref{fig:stacked_by_days} reveals that the left peak is in fact dynamically split at this epoch. The $6516\,\si{\angstrom}$ and $6456\,\si{\angstrom}$ lines at +13.9d show no sign of such splitting, but it develops in the next 4 days. Under the shock-heated circumbinary rotating disc interpretation, it is reasonable that the shock disrupts the equilibrium and introduces dynamical changes as it spreads outwards through the gas.  We used the $6516\,\si{\angstrom}$ line at +13.9d for fitting in Section \ref{sec:fitting_disc} because it reflects the pre-existing gas, not the post-shock state.

\subsection{Significance of chemistry} \label{sec:significance_chemistry}
Our observations of the disc signature are identified as being \ion{Fe}{ii} and [\ion{O}{i}] ions, and we do not observe this structure in \ion{H}{} or \ion{He}{}. Whatever the mechanism for establishing a circumbinary body of gas, if the gas was formed of recent ejecta then we would expect to see the M-shaped profile in hydrogen and helium spectral lines (\halpha$\,6563\,\si{\angstrom}$, He$\,7065\,\si{\angstrom}$, He$\,5876\,\si{\angstrom}$). We speculate that the signal is dominated by relatively heavy species because lighter particles do not remain in these orbits for long times. It is possible that, much like the Jeans escape mechanism for the atmospheres of planets, the lighter elements pick up enough kinetic energy through the temperature (via the equipartition theorem) to cease to be bound in the circumbinary disc. On the other hand, for the heavy elements, particles have less speed for the same temperature, and do not reach the required escape velocity. The notable absence of M-shaped profiles in our spectra from hydrogen or helium lines corroborates our belief that this circumbinary system pre-existed the recent explosion. This is a very preliminary argument outside of the main thrust of this work, and needs further consideration by a more detailed study.

\subsection{Circumbinary discs in classical novae}
A longstanding problem in this field is the white dwarf mass problem, detailed in \citet{Zorotovic2019}. The fundamental problem is a discrepancy whereby the theoretical model predictions of the white dwarf mass consistently underestimate the mean observed value in well-determined spectroscopic binaries of $M^{}_{WD} = 0.83\,M_{\odot}$. Consequential angular momentum loss (CAML), or angular momentum loss additional to gravitational radiation and magnetic braking, is necessary to resolve the discrepancy \citep{Schreiber2016,Liu2018}. One such mode of angular momentum loss from the inner binary could be provided by a circumbinary disc.

\citet{Spruit2000} sets out the theory behind circumbinary discs surrounding cataclysmic variables. They argue that cataclysmic variables have a high variance in the mass-transfer rate $\dot{M}$ given the small variation in white dwarf masses. A circumbinary disc applies a tidal torque to the secondary star undergoing Roche lobe overflow, increasing both angular momentum transfer and $\dot{M}$. They show that if only a small fraction from $\num{e-4}$ to $\num{e-3}$ of the total $\dot{M}$ lost from the secondary is trapped in a circumbinary disc, this disc will become massive enough on short timescales so as to influence the mass transfer. Furthermore, as this is a positive feedback mechanism, it can continue to accelerate mass transfer, which could act as the ``trigger" for a cataclysmic event such as a classical nova. It is interesting in this regard that a circumbinary disc whose existence we posit on the basis of Fig \ref{fig:best_fit_disc_uniform_density} could explain the mass problem and could act as a trigger for the nova detonation.

\subsection{Kepler argument for disc being circumbinary} \label{sec:keplerian_cb}
This section aims to compare the detected rotation speed of the disc signal with an estimate of the rotational speed of the secondary star. If the disc signal is orbiting at a slower speed than the secondary then we can conclude that the disc is orbiting on a larger radius and is therefore external to the inner binary, given the total central mass for these (assumed Keplerian) orbits is the same in both cases.

Consider the two-body problem in a circularised Keplerian system with a white dwarf, $M^{}_{\textrm{WD}}$, and a lower mass companion, $M_2 < M^{}_{\textrm{WD}}$. Defining the mass-ratio by $q = M_2/M^{}_{\textrm{WD}}$, the speed of the secondary is bounded by

\begin{equation}
    \centering
    v^{}_{2, \textrm{max}} = \frac{2 \pi}{(1+ q^{}_{\textrm{min}})} \cdot \Bigg( \frac{G}{4\pi^2} \frac{M_{\textrm{max}}}{P_{\textrm{min}}}\Bigg) ^{\frac{1}{3}}
	\label{eq:secondary_speed_upper}
\end{equation}

\begin{equation}
    \centering
    v^{}_{2, \textrm{min}} = \frac{2 \pi}{(1+ q^{}_{\textrm{max}})} \cdot \Bigg( \frac{G}{4\pi^2} \frac{M_{\textrm{min}}}{P_{\textrm{max}}}\Bigg) ^{\frac{1}{3}}.
	\label{eq:secondary_speed_lower}
\end{equation}

We then take the total binary mass to satisfy $\solarmass < M^{}_{\textrm{tot}} < 2M_{\textrm{Ch}}$ since the companion cannot in this paradigm be more massive than the WD, and the WD itself cannot be more massive than the Chandrasekhar limit $M_{\textrm{Ch}}$. We take the orbital period $81\,\si{\minute} < P < 6\,\si{\hour}$, where the lower limit comes from observations of general novae \citep{Kolb1999}. Setting the mass ratio lower limit at $q \geq 0.1$ gives $v^{}_{2, \textrm{min}} = 343\,\si{\km\per\second}$, which is comfortably larger than the $260\,\si{\km\per\second}$ rotation speed of our disc, for inclination angles $\theta < 60^\circ$. The upper bound is given by $v^{}_{2, \textrm{max}} = 783\,\si{\km\per\second}$, which is significantly larger than the detected signal. Assuming moderately low inclination angles, this is evidence that the disc we detect is at a larger orbital radius from the inner binary centre of mass than the secondary is --- in other words, it is a circumbinary disc.

The profile of the first strong detection of this feature in the $6516\,\si{\angstrom}$ \ion{Fe}{ii} line gave the outer radius as $r_{\rm out} = 14.7\,$mAU. Assuming a Keplerian inner binary, with canonical total mass $M = 1.5\,M_{\odot}$, typical orbital period from \citet{Townsley2005} for a classical nova of $P = 4\,$hr, we get a semi-major axis $a = 5.9\,$mAU for the inner binary, in the absence of any certain evidence of eccentricity. This puts the outer radius of $r_{\textrm{out}}=14.7\,$mAU outside the binary. Crucially, $r_{\textrm{out}} > 2 \times a$, which puts it in the regime of stable circumbinary orbits \citep{Doolin2011}.

\subsection{Low-density hot gas}
The same disc spectral signature is detected in the $6300\,\si{\angstrom}$ [O I] line. This is evidence that its origin is in a low-density region. It is also recently hot, as evinced by the appearance of ionised \ion{Fe}{II}. This is naturally explained in the case that it is a diffuse and expansive circumbinary disc which has just been shock-heated by a blastwave emanating from the recent explosion, so we see emission from it.

\section{Possible alternative models} \label{sec:alternatives}
There are \textit{a priori} four mechanisms for rotation in a binary system: (i) the underlying binary orbit, (ii) spin of one of the stellar components, (iii) rotation of gas within the accretion disc and (iv) rotation of gas in a family of circumbinary orbits far outside of the binary. The spectral system arises too late after the brightening to be readily explained by any of (i) to (iii). If the system were evidence of an accretion disc, it should have been seen within the first day after the first flaring episode, and the same argument applies to (ii) and (iii).  The speeds are also lower than what we would expect if they derived from inner binary activity --- as detailed in Section\,\ref{sec:keplerian_cb}.

\subsection{Bipolar outflow model}\label{sec:bipolar_model}
If this were a bipolar outflow, we would expect to see it in \halpha. We would also expect it to be on the order of ${\sim}1000\,\si{\km\per\second}$ (not a few hundred), as per \citet{Harvey2018}. Also, if it were a bipolar outflow into a decreasing density profile, you would expect it to widen in velocity space. This is an oddly transient feature, appearing at +13.9d and lasting only until +19.8d --- such transient bipolar outflows have not been observed previously and we see no reason to invoke them here.

\subsection{Pinwheel model} \label{sec:pinwheel}
We now consider the merits of a pinwheel model compared to the circumbinary disc model. In this model, instead of a reservoir of orbiting circumbinary material, some of the secondary star's wind escapes in a powerful stream of gas launched radially outward from the primary, funneled through the unstable L2 Lagrange point. As the inner binary rotates, so does this outward beam, resulting in a spiral/pinwheel shape \citep{Tuthill2007}. We use a toy model in which the system launches blobs of matter in a pinwheel, and the intensity contribution of each expanding blob decreases. Simulated spectra generated via this model show the characteristic double peaks that we observe in \novacar. 

Let us first consider the timing data, per Section \ref{sec:significance_timing}. The M-shaped signal in our observations starts up at +13.9d, and lasts with clear signal for approximately one week before fading into the low signal-to-noise regime. This is naturally explained by the circumbinary disc model, as explained in Section \ref{sec:significance_timing}. In the pinwheel model however, there is no natural geometric reason why the gas blown off the secondary would only light up two weeks after the initial peak of luminosity. Moreover, it is not clear why, having been established, the pinwheel would turn off after seven days.

The second piece of evidence is the chemistry, detailed in Section \ref{sec:significance_chemistry}. Our observations show M-shaped profiles clearly for \ion{Fe}{ii} and [\ion{O}{i}], but we do not see it in either \halpha\ or \ion{He}. For the circumbinary disc model, we reason that since we are seeing a pseudo-stable orbiting structure, formed prior to the most recent outburst event, there has been sufficient time for the lighter elements to have escaped. This is similar to the Jeans escape mechanism whereby the kinetic energy of lighter gases causes them to be unbound, and they leave a planet's atmosphere. The pinwheel model allows no time to lose the inevitable \ion{H}{} and \ion{He}{}, in disagreement with our observations.

The third piece of evidence is the spectral shape itself. In Figure \ref{fig:best_fit_disc_uniform_density}, the drop-off edge at speeds higher than the peak intensity is slanting, rather than being a vertical immediate drop-off. The disc model predicts this naturally, since there is matter closer to the centre which orbits faster than the modal velocity. Under the pinwheel model, fluid parcels are launched with a sinusoidally varying line-of-sight velocity, such that any bolus launched parallel to our line of sight has the maximum perceived speed. To fit the observations at higher speeds than the peak speed, one would need to invoke that the pinwheel engine is launching matter at a variety of speeds. This assumption is not required by the circumbinary disc model.

There is spectral profile asymmetry in some of the spectral shapes shown in Figure \ref{fig:stacked_by_days} which is not explained by a purely axisymmetric circumbinary disc. However, the simulations of \citet{Doolin2011} contend that the stability of many circumbinary orbits comes at the expense of those orbits precessing (this is especially the case for inner binaries having mass ratios that differ from unity or that have significant eccentricity).  Precession would act to tilt parts of the circumbinary disc in and out of the direct line-of-sight, readily explaining the asymmetric spectral profiles.

Occam's razor favours the circumbinary disc. Circumbinary orbits have been shown to be very stable (under certain conditions, satisfied by \novacar) by \citet{Doolin2011}, and what is stable and possible must be expected to occur in nature. There are many examples of circumbinary orbits, including Be star HR 2142 \citep{Peters2016}, SS433 \citep{Perez2010}, and 99 Herculis \citep{Kennedy2012}.  While there exist elegant manifestations of pinwheel features in nature, for example in a few Wolf-Rayet stars \citep{Tuthill2007}, we feel that a pinwheel model is neither warranted by our data, nor consistent with it.

\subsection{Outflowing disc model}
Could the M-shaped profiles be caused by an outflowing disc as opposed to a rotating disc? We argue that the fundamental spectral shape, as discussed for the pinwheel model in Section \ref{sec:pinwheel}, cannot match the observations. The edge ascent is vertical for an outflowing disc model, while only the rotating disc model can account for the steady decrease in flux at higher radial speeds than the modal speed. Despite this, it is possible that the rotating disc is also expanding radially outwards with a similar speed to the rotation. The shock arriving at the circumbinary disc lights it up, and it is possible that it also accelerates it. Any expanding disc is nonetheless likely to be circumbinary pre-existing material because of the timing reasons detailed in Section \ref{sec:significance_timing}. 

\subsection{Spectral comparison of models}
\begin{figure}
	\includegraphics[width=\columnwidth]{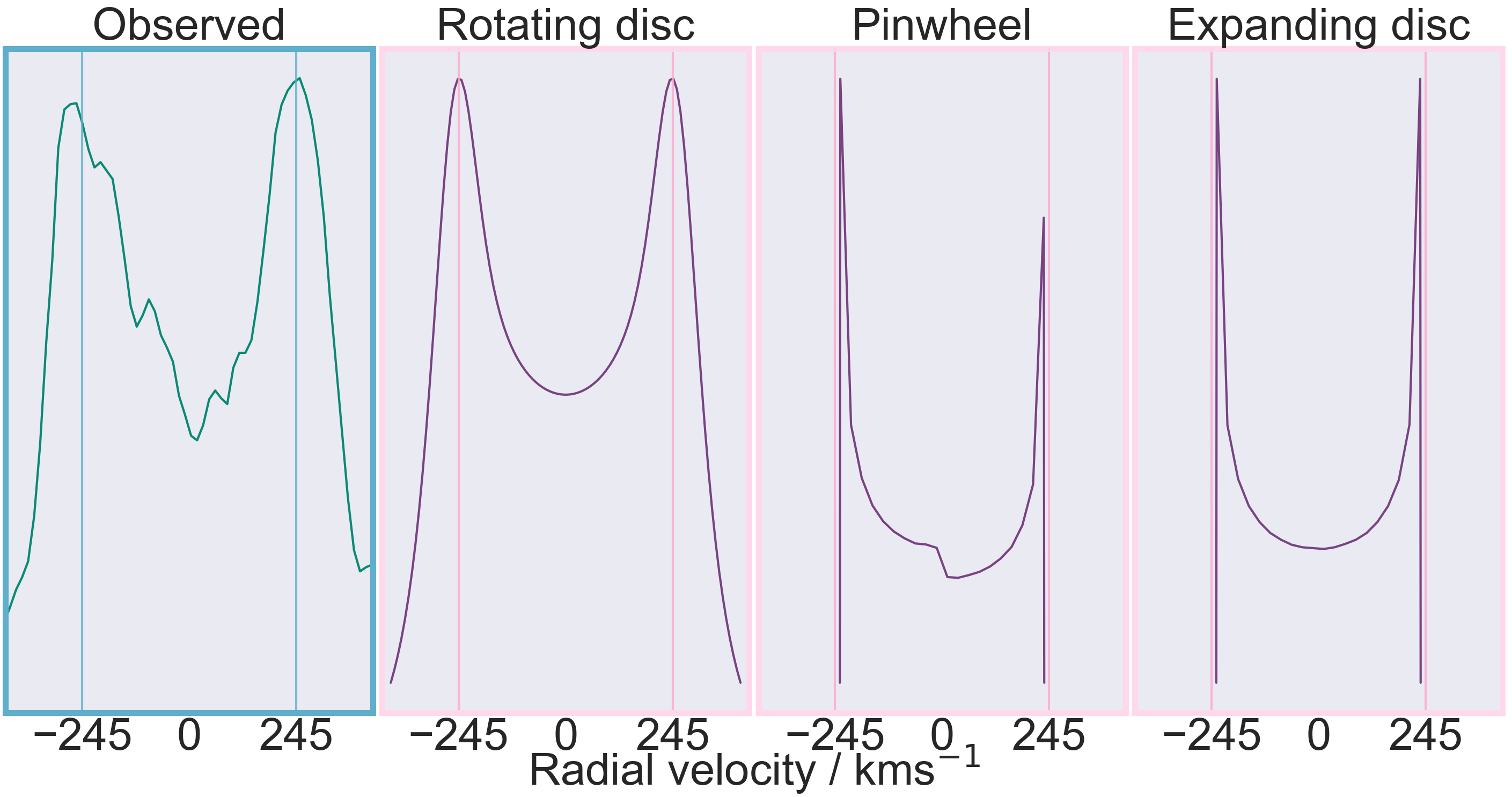}
    \caption{Data, left, compared to three models - rotating disc, pinwheel, and expanding disc. Spectral feature from the $6516\,\si{\angstrom}$ \ion{Fe}{ii} line signal, taken at +13.928d at the GJW-CL observatory in Chile.}
    \label{fig:model_comparison}
\end{figure}

We have presented four models to explain this spectral feature. We rule out the bipolar outflow model because it predicts much greater speeds than observed, together with its manifestation in \ion{H}{} and \ion{He}{}. Figure \ref{fig:model_comparison} shows a comparison between the spectral shape determined through the remaining three models --- rotating disc, pinwheel and expanding disc. Due to geometric considerations, only the rotating disc model predicts a sloping drop-off beyond the peaks. In addition, the rotating disc best reproduces the width of the peaks and is therefore our favoured model.

\section{Conclusions} \label{sec:conclusions}
We have presented observations of \ion{Fe}{ii} and [\ion{O}{i}] emission features in time-resolved spectroscopy of \novacar. We showed its consistency with being a circumbinary disc through fitting a model to the spectral shape. The possibility that the signature in fact is related to two jets was rejected due to the non-detection of this feature in \halpha\ and \ion{He}{}. The rotating circumbinary disc model fits our data better than a pinwheel scenario and two other models. We present three separate pieces of evidence --- timing data, chemistry, and spectral shape. We outlined a simple argument using orbital mechanics which gives a strong possibility that this disc is in fact circumbinary. The theoretical reasons why circumbinary discs may have a large role to play in explaining the white dwarf mass problem in close binary evolution for the classical nova scenario were noted. Hence we deduce that the M-shaped profile signal rotation is seated in circumbinary gas.

To establish the prevalence of circumbinary discs in novae is challenging, because the signal is even shorter-lived than the nova event itself. Both nova evolution studies in general, and investigations of their putative circumbinary discs in particular, necessitate sustained highly time-resolved spectroscopic observations. 

\section*{Acknowledgements}
A great many organisations and individuals have contributed to the success of
the Global Jet Watch observatories and these are listed
at \href{www.GlobalJetWatch.net}{www.GlobalJetWatch.net} but we particularly thank
the University of Oxford and the Australian Astronomical
Observatory. DM thanks the STFC for a doctoral studentship, and Oriel College, Oxford, for a graduate scholarship. We acknowledge with thanks the variable star observations from the AAVSO International Database contributed by observers worldwide and used in this research. We thank the reviewers for careful reading and insight.

%%%%%%%%%%%%%%%%%%%%%%%%%%%%%%%%%%%%%%%%%%%%%%%%%%



\bibliographystyle{mnras}
\bibliography{references} % if your bibtex file is called example.bib

%%%%%%%%%%%%%%%%%%%%%%%%%%%%%%%%%%%%%%%%%%%%%%%%%%

%%%%%%%%%%%%%%%%% APPENDICES %%%%%%%%%%%%%%%%%%%%%

% \appendix

% \section{Some extra material}

% If you want to present additional material which would interrupt the flow of the main paper,
% it can be placed in an Appendix which appears after the list of references.

%%%%%%%%%%%%%%%%%%%%%%%%%%%%%%%%%%%%%%%%%%%%%%%%%%

% Don't change these lines
\bsp	% typesetting comment
\label{lastpage}
\end{document}